\documentclass[sigconf]{acmart}
\setcopyright{none}
\renewcommand\footnotetextcopyrightpermission[1]{} % removes footnote with conference information in first column
\acmConference[SAST'25]{X Brazilian Symposium on Systematic and Automated Software Testing}{September 22–26, 2025}{Recife, PE}
\usepackage{booktabs}
\usepackage[brazil]{babel}
\usepackage[utf8]{inputenc}

\addto\captionsbrazil{%
  \renewcommand{\abstractname}{RESUMO}%
  \renewcommand{\refname}{REFERÊNCIAS}%
  \renewcommand{\keywordsname}{PALAVRAS-CHAVE}%
}
\usepackage{pifont} % para o símbolo ✔
\newcommand{\cmark}{\ding{51}} % ✔
\newcommand{\xmark}{\textemdash} % —
\usepackage{listings}
\usepackage[most]{tcolorbox}
\usepackage{xcolor}

\definecolor{systemblue}{RGB}{230,240,255}
\definecolor{usergray}{RGB}{245,245,245}
\definecolor{bgblue}{RGB}{220,235,255}   % azul clarinho
\definecolor{bgyellow}{RGB}{255,250,200} % amarelo suave
\definecolor{rq1bg}{RGB}{230,250,230}  % verde claro
\definecolor{rq2bg}{RGB}{240,222,255}  % lilás suave
\definecolor{rq3bg}{RGB}{255,245,210}  % amarelo claro

\AtBeginDocument{%
  }

\begin{document}

\title{Detecção de Conflitos Semânticos com Testes Gerados por LLM}

\author{Nathalia Barbosa}
\affiliation{%
  \institution{Centro de Informática, Universidade Federal de Pernambuco}
  \city{Recife}
  \country{Brasil}
}
\email{nfab@cin.ufpe.br}

\author{Paulo Borba}
\affiliation{%
  \institution{Centro de Informática, Universidade Federal de Pernambuco}
  \city{Recife}
  \country{Brasil}
}
\email{phmb@cin.ufpe.br}

\author{Léuson Da Silva}
\affiliation{%
  \institution{Polytechnique Montreal}
  \city{Montreal}
  \country{Canadá}
}
\email{leuson-mario-pedro.da-silva@polymtl.ca}

\begin{abstract}
Conflitos semânticos ocorrem quando um desenvolvedor introduz mudanças em uma base de código que afetam, de maneira não intencional, o comportamento de mudanças integradas em paralelo por outros desenvolvedores.
Como as ferramentas de \emph{merge} usadas na prática não conseguem detectar esse tipo de conflito, foram propostas ferramentas complementares, como SMAT, que é baseada na geração e execução de testes de unidade na linguagem Java.
Apesar de apresentar boa capacidade de detecção de conflitos, SMAT apresenta alta taxa de falsos negativos (conflitos existentes mas não sinalizados pela mesma). 
Parte desse problema, deve-se às limitações naturais de ferramentas de geração de testes de unidade, no caso, Randoop e EvoSuite. 
Para entender se essas limitações podem ser superadas por modelos de linguagem de grande porte (LLMs), este trabalho propõe, e integra ao SMAT, uma nova ferramenta de geração de testes baseada no Code Llama 70B. 
Exploramos então a capacidade desse modelo de gerar testes, com diferentes estratégias de interação, \emph{prompts} com diferentes conteúdos, e diferentes configurações de parâmetros do modelo. 
Avaliamos os resultados com duas amostras distintas, um \emph{benchmark} com sistemas mais simples, usados em trabalhos relacionados, e uma amostra mais significativa baseada em sistemas complexos e utilizados na prática. 
Por fim, avaliamos a eficácia da nova extensão do SMAT na detecção de conflitos. 
Os resultados indicam que, embora a geração de testes por LLM em cenários complexos ainda represente um desafio e seja computacionalmente custosa, há potencial promissor na identificação de conflitos semânticos.
\end{abstract}


\maketitle

\section{Introdução}

Durante o desenvolvimento colaborativo de software, conflitos podem surgir durante a integração (\textit{merge}) das alterações~\cite{zhang2022using}, especialmente quando múltiplos desenvolvedores modificam a mesma linha ou linhas consecutivas concorrentemente~\cite{10.1145/3478019}.
Esses são os chamados conflitos textuais, que exigem intervenção manual para serem resolvidos.

Embora eficazes em detectar conflitos textuais, as ferramentas de integração atuais não identificam conflitos semânticos, que ocorrem quando alterações paralelas afetam o comportamento do sistema de forma não intencional, resultando em falhas de testes ou falhas em tempo de execução mesmo após compilação bem-sucedida.
Esses \textit{conflitos semânticos comportamentais}~\cite{DASILVA2024112070} diferem de conflitos de \textit{build}~\cite{DASILVA2024112070, crystal2011, palantir2012} e podem permanecer ocultos por várias versões, tornando sua detecção um desafio.

Para detectar conflitos semânticos, o SMAT~\cite{leuson2020} utiliza geração automatizada de testes de unidade e aplica heurísticas sobre os resultados para sinalizar possíveis conflitos. No entanto, apesar de sua eficácia, o SMAT apresenta limitações, especialmente uma alta taxa de falsos negativos~\cite{DASILVA2024112070}, relacionada às capacidades limitadas de ferramentas como Randoop~\cite{pacheco2007randoop} e EvoSuite~\cite{fraser2011evosuite}, que frequentemente não cobrem todos os cenários relevantes.

Diante dos avanços em modelos de linguagem de grande porte (LLMs), surge a hipótese de que essas limitações podem ser superadas. Modelos como o Code Llama 70B~\cite{meta2023codellama} têm mostrado potencial na geração de código e testes automatizados~\cite{testpilot,aster,citywalk}. Este trabalho propõe e integra ao SMAT uma nova ferramenta de geração de testes baseada no Code Llama 70B, explorando diferentes estratégias de \textit{prompt engineering} e configurações do modelo para detectar conflitos semânticos.

Realizamos um estudo empírico com um dataset de cenários de \textit{merge}, alguns deles contendo conflitos semânticos, avaliando diferentes estratégias de \textit{prompt} (\textit{zero-shot} e \textit{1-shot}), variações de temperatura, \textit{seed} e contexto fornecido ao modelo.
Os resultados mostram que a configuração \textit{zero-shot} com temperatura 0.0 apresentou melhor desempenho no \textit{dataset} avaliado, identificando o maior número de conflitos em uma única execução, incluindo um conflito inédito não detectado em trabalhos prévios~\cite{DASILVA2024112070, leuson2020}.
Avaliamos também o impacto da complexidade dos \textit{datasets} utilizando um segundo \textit{dataset} com sistemas mais simples selecionado do trabalho de~\citet{aster}, constatando que a taxa de compilação dos testes varia significativamente conforme a complexidade do código.

Assim, as principais contribuições deste trabalho são: a análise do efeito de diferentes \textit{prompts}, variações de configuração e contextos fornecidos a um modelo de linguagem para geração de casos de teste visando detectar conflitos semânticos; a evidência de que LLMs podem gerar testes capazes de detectar conflitos semânticos em cenários de \textit{merge}, contribuindo para ampliar o potencial do SMAT em detectar conflitos; e a integração estrutural de um modelo de linguagem à arquitetura do SMAT, facilitando a adição de outros modelos.

O restante deste artigo está organizado da seguinte forma: a Seção~\ref{sec:motivacao} apresenta a motivação e o funcionamento do SMAT; a Seção~\ref{sec:metodologia} detalha a metodologia e integração da ferramenta baseada no Code Llama 70B; a Seção~\ref{sec:resultados} avalia o desempenho e discute os resultados; a Seção~\ref{sec:ameacas-a-validade} discute as ameaças à validade; a Seção~\ref{sec:trabalhos-relacionados} aborda trabalhos relacionados; e a Seção~\ref{sec:conclusao} apresenta conclusões e direções futuras.
\section{Motivação}
\label{sec:motivacao}

Conflitos semânticos geralmente passam despercebidos por ferramentas tradicionais de controle de versão, que focam apenas em diferenças textuais e ignoram interações comportamentais entre as mudanças integradas.
Para exemplificar o que são esses conflitos, considere o seguinte exemplo: dois desenvolvedores, Nivan e Renato, trabalham simultaneamente em um projeto, em um mesmo método de uma classe, conforme mostrado na Listagem~\ref{lst:cleantext}.
\begin{lstlisting}[language=Java, caption={Método alterado pelos desenvolvedores}, label={lst:cleantext}]
cleanText() {
  (*@\colorbox{bgblue}{normalizeWhiteSpace();}@*)
  removeComments();
  (*@\colorbox{bgyellow}{removeDuplicateWords();}@*)
}
\end{lstlisting}

Com o objetivo de tornar o texto mais legível, ambos fazem contribuições ao método \texttt{cleanText()} da classe \texttt{Text}.
Nivan implementa e adiciona uma chamada ao método \texttt{normalizeWhiteSpace()}, que substitui múltiplos espaços por um único espaço. Vamos chamar a versão do código com essa alteração de Left.
Renato implementa e adiciona uma chamada ao método \texttt{removeDuplicateWords()}, que remove palavras duplicadas consecutivas. Vamos chamar essa outra versão do código com tal alteração de Right.
A versão integrada do código, que combina as alterações de ambos, é chamada de Merge, e esta é a versão ilustrada na Listagem~\ref{lst:cleantext}.
Ambos testam suas funções individualmente e elas funcionam, no entanto, quando o código é integrado, um conflito semântico surge: a ordem de execução dos métodos afeta o resultado final do texto processado.
Observe, por exemplo, o comportamento da função \texttt{cleanText()} quando aplicada a uma entrada como \texttt{HELLO\textvisiblespace\textvisiblespace HELLO\textvisiblespace\textvisiblespace WORLD}.
Se a versão Left for executada, ou seja, se apenas \texttt{normalizeWhiteSpace()} for aplicada, o resultado será \texttt{HELLO\textvisiblespace HELLO\textvisiblespace WORLD}, mantendo duas ocorrências de \texttt{HELLO} e um espaço entre as palavras.
Por outro lado, se a versão Right for executada, o resultado será \texttt{HELLO\textvisiblespace\textvisiblespace\textvisiblespace\textvisiblespace WORLD}, com apenas uma ocorrência de \texttt{HELLO} e quatro espaços entre as palavras.
Quando ambas as funções são aplicadas em sequência (versão Merge), o efeito combinado revela um comportamento inesperado: \texttt{HELLO\textvisiblespace\textvisiblespace WORLD}.
Uma das ocorrências de \texttt{HELLO} é removida, mas os espaços entre as palavras permanecem, e eles só passam a constituir uma duplicação após a chamada do último método.
Esse é um exemplo clássico de conflito semântico, em que a ordem de aplicação das funções afeta o resultado final e o comportamento emergente não é trivialmente previsível.

A interação entre essas mudanças só revela seu efeito colateral quando um teste é executado com uma entrada específica (contendo tanto palavras quanto espaços duplicados), e tal entrada pode não estar presente na suíte de testes original, uma vez que os desenvolvedores tendem a criar casos de teste focados nos cenários mais comuns de uso e que validam as funcionalidades de forma isolada.
Assim, esse tipo de problema é difícil de identificar com inspeção manual, reforçando a necessidade de abordagens automatizadas baseadas em testes para detectar tais conflitos.

\subsection{SMAT}
\label{subsec:revisao_smat}

Para resolver o problema ilustrado, foi proposto o SMAT~\cite{leuson2020,DASILVA2024112070}, que detecta conflitos semânticos utilizando geração automatizada de testes de unidade em cenários de \textit{merge}, mas pode ser aprimorado por meio da integração com modelos de linguagem avançados como o Code Llama 70B~\cite{meta2023codellama}.
A ferramenta recebe como entrada cenários de \textit{merge}, cada um incluindo uma quádrupla de \textit{commits} --- Base (versão comum), Left e Right (modificações dos desenvolvedores) e Merge (integração) --- além dos caminhos dos arquivos de \textit{build}, classes-alvo e elementos alterados simultaneamente por Left e Right.

A arquitetura do SMAT é composta pelos módulos \emph{Test Generation}, \emph{Test Execution}, \emph{Test Dynamic Analysis} e \emph{Output Generation}. O funcionamento desses módulos se resume em três etapas principais:
\begin{itemize}
    \item \textbf{Geração de testes}: Emprega ferramentas como EvoSuite, Differential EvoSuite, Randoop e Randoop Clean para criar automaticamente testes para as versões Left e Right do código. A arquitetura do SMAT é extensível, permitindo a inclusão de novas ferramentas de geração de testes conforme necessário.
    \item \textbf{Compilação e execução}: As suítes de testes geradas são compiladas e executadas em todas as quatro versões do código (Base, Left, Right e Merge), com execução repetida para garantir confiabilidade dos resultados. Resultados inválidos são descartados.
    \item \textbf{Relatórios e heurísticas}: Interpreta os resultados dos testes aplicando heurísticas específicas que identificam padrões característicos de conflitos, como quando um teste falha em Base e Merge mas passa em Left ou Right, ou quando apenas Merge apresenta comportamento divergente.
\end{itemize}

Os experimentos com o SMAT demonstram a viabilidade da abordagem: inicialmente~\cite{leuson2020}, identificou corretamente 4 conflitos em 15 casos sem falsos positivos. Em trabalhos posteriores~\cite{DASILVA2024112070}, com 85 cenários e quatro ferramentas de geração, detectou 9 conflitos entre 29, mas com três falsos positivos e uma taxa considerável de falsos negativos.

Apesar dos avanços, ainda existem limitações, sobretudo relacionadas à cobertura dos testes automatizados, que podem não identificar interações sutis entre as versões Left e Right, levando aos falsos negativos mencionados.
Essas interações incluem dependências indiretas ou efeitos colaterais não explícitos, que podem passar despercebidos caso os testes gerados não explorem o caminho que revela o conflito.
Ferramentas tradicionais frequentemente não capturam essas nuances, tornando a integração de LLMs uma oportunidade de explorar novas estratégias de geração de testes para potencializar a detecção de conflitos semânticos.

A incorporação do Code Llama 70B~\cite{meta2023codellama} ao SMAT permite avaliar e aprimorar métodos de geração de testes e detecção de conflitos.
A escolha deste LLM foi fundamentada em sua especialização em código (incluindo Java), seu equilíbrio entre capacidade de raciocínio e viabilidade computacional, e sua disponibilidade open-source, permitindo reprodutibilidade e controle sobre o ambiente de execução.
\section{Metodologia}
\label{sec:metodologia}

Para investigar se LLMs podem ser úteis para geração de testes de unidade com o objetivo de detectar conflitos semânticos, conduzimos uma série de experimentos utilizando duas amostras distintas, cada uma com motivações e características específicas.
O primeiro \textit{dataset} contém cenários reais de \textit{merge} com conflitos semânticos, permitindo avaliar a capacidade do modelo em detectar esse tipo de problema.
Já o segundo \textit{dataset} não contém cenários de \emph{merge} nem conflitos semânticos, e é composto por versões finais de projetos mais simples, sendo utilizado para avaliar a capacidade de geração e compilação de testes pelo modelo.
Para viabilizar esses experimentos, desenvolvemos e integramos ao SMAT um novo módulo de geração de testes automatizados, cujo funcionamento é baseado no modelo Code Llama 70B.
A seguir, o processo de integração do módulo ao SMAT é apresentado, bem como os \textit{datasets} selecionados para o estudo e informações adicionais sobre os procedimentos adotados durante os experimentos.

\subsection{Integração do Módulo Code Llama 70B}

Como mencionado na Seção~\ref{sec:motivacao}, a arquitetura do SMAT é modular, permitindo a adição de novas ferramentas e funcionalidades conforme necessário.
Neste contexto, desenvolvemos um módulo completo de geração de testes automatizados, que utiliza o modelo Code Llama 70B como parte central, mas incorpora diversas etapas adicionais de processamento, integração e validação. O novo módulo expande o \emph{Test Generation} já existente, agregando um pipeline que realiza desde a extração e preparação das informações do código-fonte, passando pela construção e envio de prompts ao modelo, até o tratamento, validação e integração dos testes gerados ao fluxo do SMAT.

\paragraph{Visão Geral da Integração}

\begin{figure*}[ht]
    \centering
    \includegraphics[width=\textwidth]{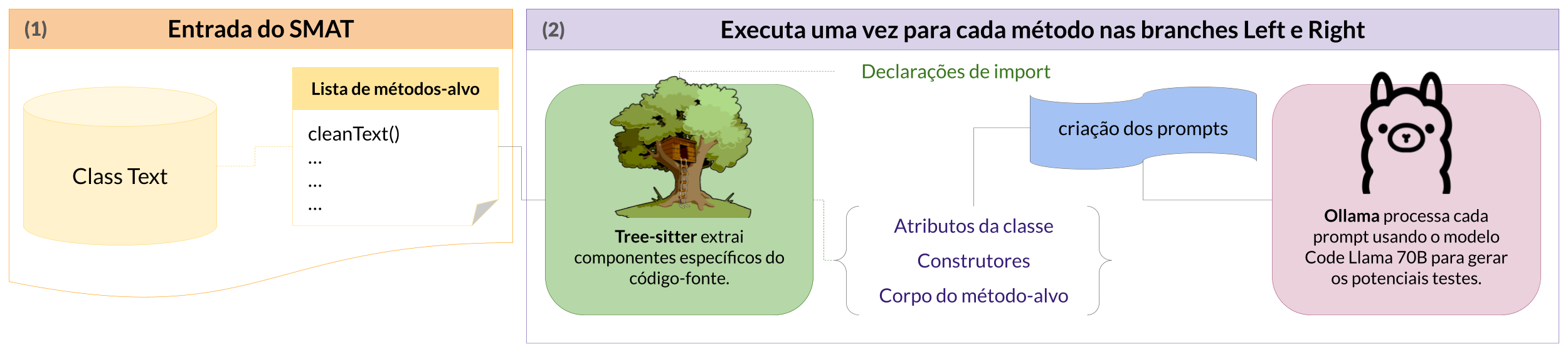}
    \caption{Arquitetura da ferramenta baseada no Code Llama 70B integrada ao SMAT.}
    \Description{Diagrama mostrando a integração entre a ferramenta proposta e o SMAT, incluindo etapas como extração de informações, construção de \textit{prompts} e invocação do modelo.}
    \label{fig:smat-architecture}
\end{figure*}

A integração da nova ferramenta ao SMAT é implementada através de um \textit{pipeline} que utiliza informações do código em análise para gerar testes de unidade executáveis.
O fluxo inicia com extração de elementos relevantes do código-fonte, que são então formatados em \textit{prompts} estruturados para serem enviados ao LLM.
A resposta do modelo contendo o código dos testes passa por um processo de limpeza e validação antes de ser compilado e executado pelos demais componentes do SMAT.
Este processo é organizado em etapas sequenciais, conforme detalhado a seguir:

Além dos elementos citados na Seção~\ref{subsec:revisao_smat}, a entrada do SMAT foi alterada para receber, junto a cada elemento alterado simultaneamente por Left e Right na classe-alvo, um resumo textual das mudanças realizadas por cada parente. Essa mudança foi incluída para avaliar se essa informação ajudaria ou não o LLM a gerar testes que detectam conflitos.
Conforme ilustrado na Figura~\ref{fig:smat-architecture}, o passo (1) mostra as informações de entrada que são inicialmente usadas para a geração dos \textit{prompts}: as classes-alvo (\texttt{Class Text}) e os seus respectivos métodos-alvo, que são os elementos alterados por Left e Right os quais serão testados (\texttt{cleanText()}).
Como destacado no passo (2), para cada um dos métodos-alvo informados na entrada, utilizamos a biblioteca de parsing \emph{Tree-sitter}~\cite{treesitter} para construir a AST (\textit{Abstract Syntax Tree}) do código-fonte, o que permite estruturar o código de forma a coletar informações para a geração dos \textit{prompts}.
A partir dessa AST, extraímos atributos da classe, construtores e o corpo do método, além das declarações de \texttt{import}, que são anexadas aos testes no fim da iteração para garantir sua compilação.
Com poucas modificações, essa biblioteca pode ser utilizada para gerar ASTs de qualquer linguagem de programação, o que torna a ferramenta proposta adaptável a outros contextos além do Java.

\paragraph{Construção dos prompts}

As informações extraídas são organizadas em diferentes formatos de \textit{prompts}, compostos por mensagens dos tipos \texttt{system}, \texttt{user} e \texttt{assistant}, cada uma com uma função específica na interação com o modelo de linguagem.
Segundo~\citet{white2023promptpatterncatalogenhance}, os \textit{prompts} podem ser estruturados por meio de \textit{prompt patterns}, que funcionam como padrões reutilizáveis para resolver problemas recorrentes no uso de LLMs. 
A mensagem do tipo \texttt{system} corresponde ao padrão \textit{Persona}, atribuindo ao modelo uma identidade ou papel específico, com o objetivo de orientar seu comportamento ao longo da conversa. Essa mensagem estabelece o contexto geral da interação e define diretrizes para o estilo e o conteúdo das respostas geradas. Abaixo está a mensagem do tipo \texttt{system} utilizada neste estudo:

\begin{tcolorbox}[colback=systemblue, colframe=blue!30!black, title=SYSTEM, breakable]
You are a senior Java developer with expertise in JUnit testing.\\
Your task is to provide JUnit tests for the given method in the class under test, considering the changes introduced in the left and right branches.\\
You have to answer with the test code only, inside code blocks (\verb|```|).\\
The tests should start with \verb|@Test|.
\end{tcolorbox}

As mensagens do tipo \texttt{user} fornecem a entrada principal ao modelo, podendo ser relacionadas aos padrões \textit{Recipe} e \textit{Context Manager}. Nessas mensagens estão descritas informações do código, o método-alvo a ser testado e instruções para o modelo, delimitando o escopo da tarefa e os requisitos desejados. A seguir, um exemplo de mensagem do tipo \texttt{user} que nosso módulo usa para gerar testes para o método \texttt{cleanText} da classe \texttt{Text}:
\begin{tcolorbox}[colback=usergray, colframe=gray!50!black, title=USER, breakable]
Here is the context of the method under test in the class \texttt{Text} on the \texttt{left} branch:

\medskip

\textbf{Class fields:}
\begin{quote}\ttfamily
public String text;
\end{quote}

\textbf{Constructors:}
\begin{quote}\ttfamily
public Text(String text) \{\\
\ \ this.text = text;\\
\}
\end{quote}

\textbf{Target Method Under Test:}
\begin{quote}\ttfamily
public void cleanText() \{\\
\ \ Text inst = new Text(text);\\
\ \ inst.normalizeWhiteSpace();\\
\ \ inst.removeComments();\\
\ \ this.text = inst.text;\\
\}
\end{quote}

\medskip

Now generate JUnit tests for the method under test, considering the given context. Remember to create meaningful assertions.\\
Write all tests inside code blocks (\verb|```|), and start each test with \verb|@Test|.
\end{tcolorbox}

Nessa mensagem, que compõe um \textit{prompt} do tipo \textit{zero-shot}, o contexto do método-alvo é fornecido, incluindo os atributos da classe, construtores e o corpo do método. Então, o modelo é instruído a gerar testes JUnit para esse método, considerando as informações fornecidas.

Já as mensagens do tipo \texttt{assistant} são utilizadas aqui apenas na configuração \textit{1-shot}. Esse uso está alinhado ao padrão \textit{Template}, cujo objetivo é fornecer um exemplo que sirva de modelo para a geração de saídas consistentes e adequadas à tarefa proposta.

Os \textit{prompts} utilizados foram construídos a partir de conjuntos de mensagens com diferentes partes do contexto dos cenários: (i) resumo das mudanças introduzidas por Left e Right, (ii) atributos da classe, (iii) construtores e (iv) corpo do método-alvo.
Para cada formato (\textit{zero-shot} e \textit{1-shot}), conforme está listado na Tabela~\ref{tab:conflitos_por_prompt}, existem 8 variações de \textit{prompt}, cada uma combinando diferentes subconjuntos dessas partes.
Isso foi feito para explorar quais informações são mais relevantes para a geração de testes eficazes, trazendo mais variedade nas interações com o modelo e, assim, aumentando a diversidade dos testes gerados.

\begin{itemize}
    \item \textbf{Zero-shot:} O prompt é composto por uma mensagem de sistema (\texttt{system}), seguida pelas mensagens do usuário (\texttt{user}) que fornecem o contexto do método-alvo a ser testado.

    \item \textbf{1-shot:} Neste formato, o prompt primeiro apresenta um exemplo completo de interação antes da solicitação real. A estrutura contém: a mensagem de sistema (\texttt{system}), seguida por um par de mensagens de exemplo (uma do \texttt{user} com um método fictício e uma resposta correspondente do \texttt{assistant} com o teste gerado). Somente após esse exemplo, é apresentada a mensagem do \texttt{user} com o método-alvo real.
\end{itemize}

Todos os prompts usados nos experimentos estão disponíveis no apêndice online deste trabalho~\cite{online-appendix}.

\paragraph{Invocação do modelo}

Esses \textit{prompts} são enviados, por meio da API REST do \emph{Ollama}~\cite{ollama} --- plataforma \textit{open-source} para execução local de LLMs ---, ao modelo Code Llama 70B, que retorna múltiplas respostas (em quantidade configurável) para cada método testado.
Para os experimentos realizados, apenas uma resposta foi solicitada para cada \textit{prompt}, por método-alvo em cada \textit{branch} (Left e Right).

\paragraph{Processamento da saída}

\begin{figure*}[ht]
    \centering
    \includegraphics[width=0.8\textwidth]{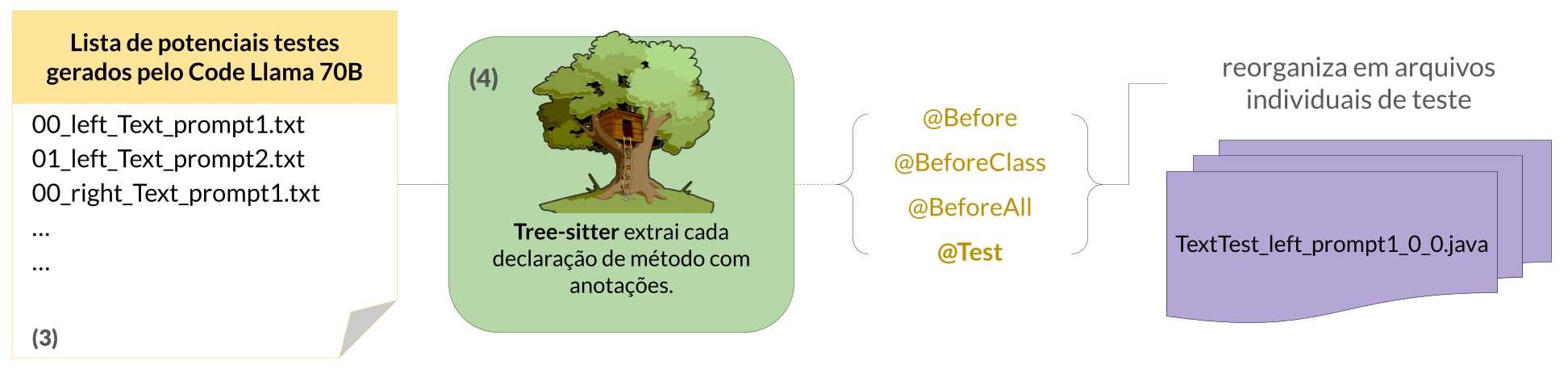}
    \caption{Processamento da saída da ferramenta baseada no Code Llama 70B.}
    \Description{Diagrama ilustrando as etapas de processamento da saída gerada pela ferramenta proposta.}
    \label{fig:smat-architecture2}
\end{figure*}

A Figura~\ref{fig:smat-architecture2} ilustra o processamento da saída da ferramenta baseada no Code Llama 70B, que ocorre após a invocação do modelo.
Conforme mostrado no passo (3), as respostas do LLM são consideradas aqui como potenciais suítes de testes.
Cada uma delas passa por uma etapa de limpeza utilizando expressões regulares, com o objetivo de remover trechos que não correspondem a código Java válido.
Isso inclui linguagem natural, comentários irrelevantes, artefatos de formatação e caracteres que não pertencem ao léxico da linguagem, os quais poderiam comprometer a compilação e execução dos testes.

Após essa limpeza, utilizamos novamente o \emph{Tree-sitter}, destacado no passo (4), para analisar a saída resultante e extrair os métodos de teste por meio da identificação de métodos anotados com a anotação \texttt{@Test}.
Isso é um primeiro passo para garantir que os testes gerados sejam válidos, pois essa anotação é essencial para que o JUnit reconheça os métodos como testes a serem executados.

Então, cada teste é armazenado em um arquivo único, identificado por um nome que concatena a classe-alvo, a \textit{branch}, o \textit{prompt}, o índice do método e um número sequencial, como em \texttt{TextTest\_left\_prompt1\_0\_0.java}.
O objetivo disso é maximizar o aproveitamento dos testes gerados: caso um ou mais testes apresentem problemas de compilação, os demais ainda poderão ser executados sem prejuízo ao processo de avaliação.
Essa abordagem também facilita a organização dos experimentos e a análise posterior dos resultados, permitindo isolar falhas e identificar padrões nos erros gerados.

\subsection{Datasets Utilizados}

A priori, utilizamos como base o \textit{mergedataset}, conforme descrito por~\citet{DASILVA2024112070}, que originalmente contém 85 cenários. Contudo, nosso conjunto de dados final foi reduzido para 79 cenários, pois 6 deles foram descartados por falta de arquivos no repositório (dois do projeto \textit{libgdx}, três do \textit{elasticsearch} e um do \textit{ReactiveX}).
Esse subconjunto do \textit{mergedataset} é composto por 29 projetos Java de código aberto, abrangendo diversas áreas de software, e inclui 29 cenários com conflitos semânticos e 50 sem conflitos.

Para complementar a análise, foram selecionados dois projetos adicionais, considerados ``toy'', que são mais simples e foram extraídos do \textit{dataset} utilizado no trabalho de~\citet{aster}. Por se tratar de um trabalho com foco apenas na geração de testes, esses projetos não contêm cenários de \textit{merge} e, portanto, não apresentam conflitos semânticos: o \textbf{Eclipse Cargo Tracker}, uma aplicação open-source de rastreamento de cargas construída com Jakarta EE e estruturada com base nos princípios de Domain-Driven Design, que simula fluxos comuns de sistemas corporativos, como monitoramento logístico, roteamento e controle de eventos~\cite{cargotracker}; e o \textbf{DayTrader 8}, um benchmark de sistema de negociação de ações desenvolvido com Java EE 8, que implementa funcionalidades típicas de sistemas financeiros, como login de usuários, consulta de contas e execução de ordens de compra e venda, sendo amplamente utilizado como referência para testes funcionais e de desempenho em servidores de aplicação Java EE~\cite{daytrader}.

Esses dois foram selecionados por simularem sistemas com comportamento próximo ao de aplicações reais, ainda que com menor complexidade estrutural e de código em comparação ao \textit{mergedataset}. Essa escolha permite avaliar a capacidade do modelo em gerar testes de unidade tanto em contextos mais controlados e didáticos quanto em cenários extraídos de sistemas reais.

\subsection{Procedimentos Experimentais}

Neste estudo, conduzimos dois experimentos distintos para responder às seguintes questões de pesquisa:

\begin{itemize}
    \item \textbf{RQ1:} Entre as configurações de \textit{prompt} e parâmetros avaliadas, alguma delas permite ao modelo gerar testes mais eficazes para identificar conflitos semânticos?
    \item \textbf{RQ2:} A ferramenta baseada no Code Llama 70B é útil para a detecção automática de conflitos semânticos em cenários de \textit{merge}?
    \item \textbf{RQ3:} Como a complexidade do código em análise impacta os resultados do LLM?
\end{itemize}

\paragraph{Configuração dos Parâmetros}

Os parâmetros do modelo explorados foram: (i) \textbf{temperatura} (0.0 para respostas determinísticas e 0.7 para maior diversidade), (ii) \textbf{\textit{seed}} (42 e 123 para avaliar estabilidade dos resultados), e (iii) \textbf{formato do \textit{prompt}} (\textit{zero-shot} e \textit{1-shot}).

\paragraph{Experimento 1: Detecção de Conflitos Semânticos}

Utilizando o \textit{mergedataset} (79 cenários), avaliamos a eficácia na detecção de conflitos semânticos (RQ1 e RQ2). Executamos 8 configurações distintas, combinando temperaturas, \textit{seeds} e formatos de \textit{prompt}, totalizando 1264 chamadas do modelo a cada execução.
Cada configuração aplica 8 \textit{prompts} distintos para cada método-alvo em cada \textit{branch}.

\paragraph{Experimento 2: Capacidade de Geração e Compilação}

Para avaliar o impacto da complexidade do código (RQ3), comparamos a geração e compilação de testes nos dois \textit{datasets}, fixando temperatura em 0.0 e aplicando 5 \textit{prompts} distintos (4 \textit{zero-shot} e 1 \textit{1-shot}). O \textit{ASTER dataset} não inclui resumos de mudanças por não conter cenários de \textit{merge}.

\paragraph{Análise dos Resultados}

As análises foram realizadas a partir dos relatórios do SMAT, incluindo métricas de tempo de execução, quantidade de testes gerados, compilados e executados, além dos conflitos detectados. Adicionalmente, verificamos manualmente os testes que identificaram conflitos semânticos para validar os resultados e evitar falsos positivos.

Todos os \textit{prompts} utilizados estão disponíveis no apêndice online deste trabalho~\cite{online-appendix}. Os experimentos foram conduzidos em uma máquina \textit{x86\_64} rodando Ubuntu, equipada com 251\,GB de RAM, processador Intel Xeon Silver 4316 (40 núcleos) e duas GPUs NVIDIA A100 80GB PCIe.
\section{Resultados}
\label{sec:resultados}

Para responder às perguntas de pesquisa discutidas na seção anterior, realizamos os experimentos descritos anteriormente e apresentamos os resultados a seguir. A seção está estruturada em torno das perguntas de pesquisa, começando com uma análise quantitativa da geração de testes, seguida pela detecção de conflitos semânticos e comparação com outras ferramentas, e finalizando com uma análise do impacto do \textit{dataset} e do tempo de execução.

\subsection{Análise Quantitativa da Geração de Testes}

\begin{table*}[h!]
\centering
\small
\renewcommand{\arraystretch}{1.2}
\begin{tabular}{@{}lccccc@{}}\toprule
\textbf{Métrica} & \multicolumn{2}{c}{\textbf{Temperatura 0}} & \phantom{ab} & \multicolumn{2}{c}{\textbf{Temperatura 0.7}} \\
\cmidrule{2-3} \cmidrule{5-6}
& \textbf{1S} & \textbf{ZS} & & \textbf{1S} & \textbf{ZS} \\
\midrule
Média de testes gerados & 10404 & 8592 & & 1383 & 2358 \\
Média de testes que compilaram & 1253 & 586 & & 124 & 121 \\
Média de testes gerados por método & 131.70 & 108.76 & & 17.50 & 29.85 \\
Média de testes que compilaram por método & 15.86 & 7.42 & & 1.57 & 1.53 \\
Taxa de compilação média & 12.04\% & 6.82\% & & 8.97\% & 5.13\% \\
Conflitos detectados & 1* & 3* & & 3* & 1* \\
\bottomrule
\end{tabular}
\caption{%
Resultados da geração de testes com a ferramenta baseada no Code Llama 70B para diferentes configurações (1S = \emph{1-shot}, ZS = \emph{zero-shot}). Valores são médias de duas execuções independentes (seeds 42 e 123). {*} indica conflito não identificado previamente por ferramentas do SMAT.
}
\label{tab:resultados_code_llama}
\end{table*}

A Tabela~\ref{tab:resultados_code_llama} apresenta os resultados da geração de testes e detecção de conflitos utilizando a ferramenta baseada no Code Llama 70B.
Para a construção da tabela, foram realizadas duas execuções para cada formato e temperatura, cada uma com 8 \textit{prompts}, e então os resultados foram agregados para obter as médias apresentadas.
A quantidade de conflitos detectados foi considerada como a união dos conflitos identificados em ambas as execuções.
Podemos observar que a temperatura 0.0 resultou em uma geração significativamente maior de testes, com uma média de 10404 testes na configuração \textit{1-shot} (1S) e 8592 na configuração \textit{zero-shot} (ZS).
Em contraste, a temperatura 0.7 gerou uma quantidade consideravelmente menor de testes, com médias de 1383 e 2358 testes para as configurações 1S e ZS, respectivamente.

A taxa de compilação também foi superior com a temperatura 0.0, atingindo o pico de 12.04\% na configuração 1S. Embora a configuração ZS tenha gerado um grande volume de testes, sua taxa de compilação foi menor (6.82\%), indicando que a presença de um exemplo (1-shot) favorece a geração de código sintaticamente mais correto.

Com temperatura 0.7, as taxas de compilação foram de 8.97\% (1S) e 5.13\% (ZS).
Todas as quatro configurações experimentais foram capazes de detectar conflitos semânticos, com cada uma identificando, pelo menos, o conflito previamente não detectado por outras ferramentas do SMAT.
Em especial, a configuração \textit{zero-shot} com temperatura 0.0 identificou três conflitos numa única execução, enquanto cada \textit{seed} da configuração \textit{1-shot} com temperatura 0.7 detectou dois, sendo um deles comum às duas \textit{seeds}.
Portanto, a configuração \textit{zero-shot} com temperatura 0.0 se destacou por detectar o maior número de conflitos semânticos em uma única execução.

\subsection{RQ1: Alguma configuração permite gerar testes mais eficazes para identificar conflitos semânticos?}

\begin{table*}[!htbp]
\centering
\small
\renewcommand{\arraystretch}{1.2}
\setlength{\tabcolsep}{4pt}
\begin{tabular}{@{}lcccccccc@{}}
\toprule
\textbf{Conflito (Projeto::Classe::Elemento)} &
\rotatebox{60}{ZS (0, 42)} &
\rotatebox{60}{ZS (0, 123)} &
\rotatebox{60}{ZS (0.7, 42)} &
\rotatebox{60}{ZS (0.7, 123)} &
\rotatebox{60}{1S (0, 42)} &
\rotatebox{60}{1S (0, 123)} &
\rotatebox{60}{1S (0.7, 42)} &
\rotatebox{60}{1S (0.7, 123)} \\
\midrule
\texttt{antlr4::Python2Target::python2Keywords}            & \cmark & \cmark & \xmark & \xmark & \xmark & \xmark & \xmark & \xmark \\
\texttt{antlr4::Python3Target::python3Keywords}            & \cmark & \cmark & \xmark & \xmark & \xmark & \xmark & \xmark & \xmark \\
\textbf{\texttt{cloud-slang::SlangImpl::getAllEventTypes()}} & \cmark & \cmark & \cmark & \cmark & \cmark & \cmark & \cmark & \cmark \\
\texttt{spring-boot::AtomikosProperties::asProperties()}   & \xmark & \xmark & \xmark & \xmark & \xmark & \xmark & \cmark & \xmark \\
\texttt{spring-boot::AtomikosPropertiesTest::testProperties()} & \xmark & \xmark & \xmark & \xmark & \xmark & \xmark & \xmark & \cmark \\
\bottomrule
\end{tabular}
\caption{%
Conflitos semânticos detectados por configuração (1S = \emph{1-shot}, ZS = \emph{zero-shot}). Valores entre parênteses: (temperatura, \textit{seed}). \cmark indica conflito detectado. Em negrito: conflito não identificado previamente por ferramentas do SMAT.
}
\label{tab:conflitos_matriz_completa}
\end{table*}

A Tabela~\ref{tab:conflitos_matriz_completa} oferece uma visão detalhada da capacidade de detecção de conflitos, evidenciando a \textbf{complementaridade entre as configurações experimentais}. Nenhuma configuração isoladamente foi capaz de detectar todos os conflitos.

O conflito \texttt{getAllEventTypes()} foi o único identificado em todas as oito execuções experimentais, além de ser o único não previamente detectado por outras ferramentas do SMAT.
Nesse cenário, tanto a \textit{branch} Left quanto a \textit{branch} Right adicionaram elementos distintos à mesma lista, resultando em tamanhos diferentes (23 vs 21 elementos).
O modelo conseguiu identificar essa inconsistência gerando testes que verificam o tamanho da lista retornada.
A Listagem~\ref{lst:exemplo_getAllEventTypes} apresenta um exemplo representativo dos testes gerados que conseguiram detectar este conflito.
O teste assume a expectativa da \textit{branch} Left (23 elementos) e, por essa razão, falha quando executado no contexto da \textit{branch} Right, revelando assim o conflito semântico.

\begin{lstlisting}[language=java, caption={Exemplo de teste gerado pelo modelo Code Llama 70B para detectar o conflito \texttt{getAllEventTypes()}.}, label={lst:exemplo_getAllEventTypes}]
@Test
public void test00() {
    SlangImpl slang = new SlangImpl();
    Set<String> eventTypes = slang.getAllEventTypes();
    assertEquals(23, eventTypes.size());
}
\end{lstlisting}

Os dois conflitos relacionados ao projeto \texttt{antlr4} foram observados apenas nas execuções \textit{zero-shot} com temperatura 0.0.
Já o conflito \texttt{asProperties()} ocorreu exclusivamente na configuração \textit{1-shot} com temperatura 0.7 e \textit{seed} 42, enquanto \texttt{testProperties()} foi detectado apenas com temperatura 0.7 e \textit{seed} 123.

Observamos uma clara especialização. As execuções em \textbf{\textit{zero-shot} com temperatura 0.0} foram as únicas capazes de identificar os dois conflitos no projeto \texttt{antlr4}.
Por outro lado, os conflitos no projeto \texttt{spring-boot} foram detectados exclusivamente pelas execuções em \textbf{\textit{1-shot} com temperatura 0.7}.

Este resultado sugere que a variação na temperatura e na estratégia de \textit{prompt} permite explorar diferentes tipos de falhas, indicando que uma abordagem combinando múltiplas configurações é mais eficaz para maximizar a detecção de conflitos.
No entanto, é importante destacar que essa estratégia é computacionalmente mais custosa.
Por exemplo, os tempos de geração de testes para cada configuração variaram de cerca de 6 horas (\textit{1-shot}, temperatura 0.7) até quase 28 horas (\textit{1-shot}, temperatura 0) para o \textit{mergedataset} inteiro.
Sendo assim, usar mais de uma configuração para detectar conflitos pode ser inviável em cenários onde o tempo é um fator crítico.
Uma descrição detalhada dos tempos de execução pode ser consultada na Seção~\ref{sec:tempo_execucao}.

\begin{table*}[!htbp]
\centering
\small
\renewcommand{\arraystretch}{1.2}
\begin{tabular}{@{}l l l@{}}
\toprule
\textbf{Prompt} & \textbf{Contexto} & \textbf{Conflitos Detectados} \\
\midrule
\texttt{prompt1} & Target method body & 1* \\
\texttt{prompt2} & Left and Right changes summary, target method body & 2* \\
\texttt{prompt3} & Class fields, target method body & 1* \\
\texttt{prompt4} & Constructors, target method body & 1* \\
\texttt{prompt5} & Left and Right changes summary, class fields, target method body & 2 \\
\texttt{prompt6} & Left and Right changes summary, constructors, target method body & 1* \\
\texttt{prompt7} & Class fields, constructors, target method body & 1 \\
\texttt{prompt8} & Left and Right changes summary, class fields, constructors, target method body & 1* \\
\bottomrule
\end{tabular}
\caption{%
Conflitos semânticos detectados por contexto de \textit{prompt}. {*} indica conflito não identificado previamente por ferramentas do SMAT.
}
\label{tab:conflitos_por_prompt}
\end{table*}

A Tabela~\ref{tab:conflitos_por_prompt} explora como diferentes contextos de \textit{prompt} influenciam a detecção. Todos os \textit{prompts} foram capazes de identificar ao menos um conflito. Destacam-se os \textit{prompts} \texttt{prompt2} e \texttt{prompt5}, que detectaram dois conflitos semânticos cada. Ambos têm como contexto o resumo das mudanças (\textit{Left and Right changes summary}); no entanto, esse fato não implica necessariamente uma relação causal entre o contexto do \textit{prompt} e a quantidade de conflitos detectados.

Essas informações nos levam a responder a primeira pergunta de pesquisa:

\begin{tcolorbox}[colback=rq1bg, colframe=green!40!black, title=RQ1]
Os resultados indicam que diferentes combinações de \textit{prompt} e parâmetros do modelo levaram à detecção de conflitos distintos. Entre as configurações avaliadas, a versão \textit{zero-shot} com temperatura 0.0 foi a mais eficaz em detectar conflitos semânticos, identificando até três conflitos em uma única execução.
\end{tcolorbox}

\begin{table}[!htbp]
\centering
\small
\renewcommand{\arraystretch}{1.2}
\begin{tabular}{@{}p{5.5cm}l@{}}
\toprule
\textbf{Ferramenta de Geração de Testes} & \textbf{Conflitos Detectados} \\
\midrule
Differential EvoSuite & 6 \\
EvoSuite & 5 [1] \\
\textbf{Code Llama 70B (união de todas as configurações experimentais)} & 5 (1*) \\
\textbf{Code Llama 70B (\textit{zero-shot}, temperatura 0.0)} & 3 (1*) \\
Randoop & 2 \\
Randoop Clean & 2 [1] \\
\bottomrule
\end{tabular}
\caption{%
Conflitos detectados por ferramenta. (*) indica conflitos detectados exclusivamente pela ferramenta. [1] representa falsos positivos. Dados extraídos de~\citet{DASILVA2024112070}.
}
\label{tab:conflitos_detectados}
\end{table}

\subsection{RQ2: A ferramenta baseada no Code Llama 70B é útil para a detecção automática de conflitos semânticos em cenários de merge?}

A Tabela~\ref{tab:conflitos_detectados} posiciona a ferramenta proposta em relação a outras ferramentas de geração de testes. A união de todas as execuções do Code Llama detectou um total de \textbf{5 conflitos}, um desempenho comparável ao da EvoSuite e superior ao da Randoop.
O destaque principal é a detecção de um \textbf{conflito inédito (*)}, que não foi identificado por nenhuma outra ferramenta, nem mesmo pela Differential EvoSuite, que detectou o maior número de conflitos --- 6, sendo 4 deles comuns aos detectados pela nossa abordagem.
Essa complementaridade sugere que as diferentes estratégias de geração de testes exploram aspectos distintos do código, indicando que uma abordagem híbrida combinando as ferramentas poderia maximizar a detecção de conflitos semânticos.
Além disso, a melhor configuração individual (\textit{zero-shot}, temperatura 0.0) foi capaz de encontrar 3 conflitos, superando ferramentas estabelecidas como a Randoop e a Randoop Clean, que detectaram apenas 2 conflitos cada.
Os tempos aproximados para a execução dessas configurações foram de cerca de 1 dia para a configuração \textit{zero-shot} com temperatura 0.0 e mais de 5 dias para a união de todas as configurações experimentais.
Um ponto negativo, portanto, é que a ferramenta proposta é substancialmente mais lenta do que as demais.
É importante notar que, ao contrário de ferramentas como EvoSuite e Randoop Clean, não foram registrados falsos positivos nos resultados da integração da ferramenta baseada no Code Llama 70B.
A detecção de conflitos não depende apenas de testes que falham ou passam, mas sim das heurísticas específicas do SMAT que analisam padrões de comportamento divergente entre as versões Base, Left, Right e Merge, mitigando o risco de falsos positivos devido a testes triviais.
Com isso, podemos responder à segunda pergunta de pesquisa:

\begin{tcolorbox}[colback=rq2bg, colframe=purple!40!black, title=RQ2]
A ferramenta baseada no modelo Code Llama 70B demonstrou ser eficaz para a detecção automática de conflitos semânticos em cenários de \textit{merge}, superando ferramentas tradicionais como Randoop e EvoSuite em termos de conflitos detectados. Além disso, foi capaz de identificar um conflito inédito não detectado por outras ferramentas do SMAT, embora sua execução tenha sido significativamente mais lenta.
\end{tcolorbox}

\subsection{RQ3: Como a complexidade do código em análise impacta os resultados do LLM?}

\begin{figure}[htbp]
    \centering
    \includegraphics[width=\columnwidth]{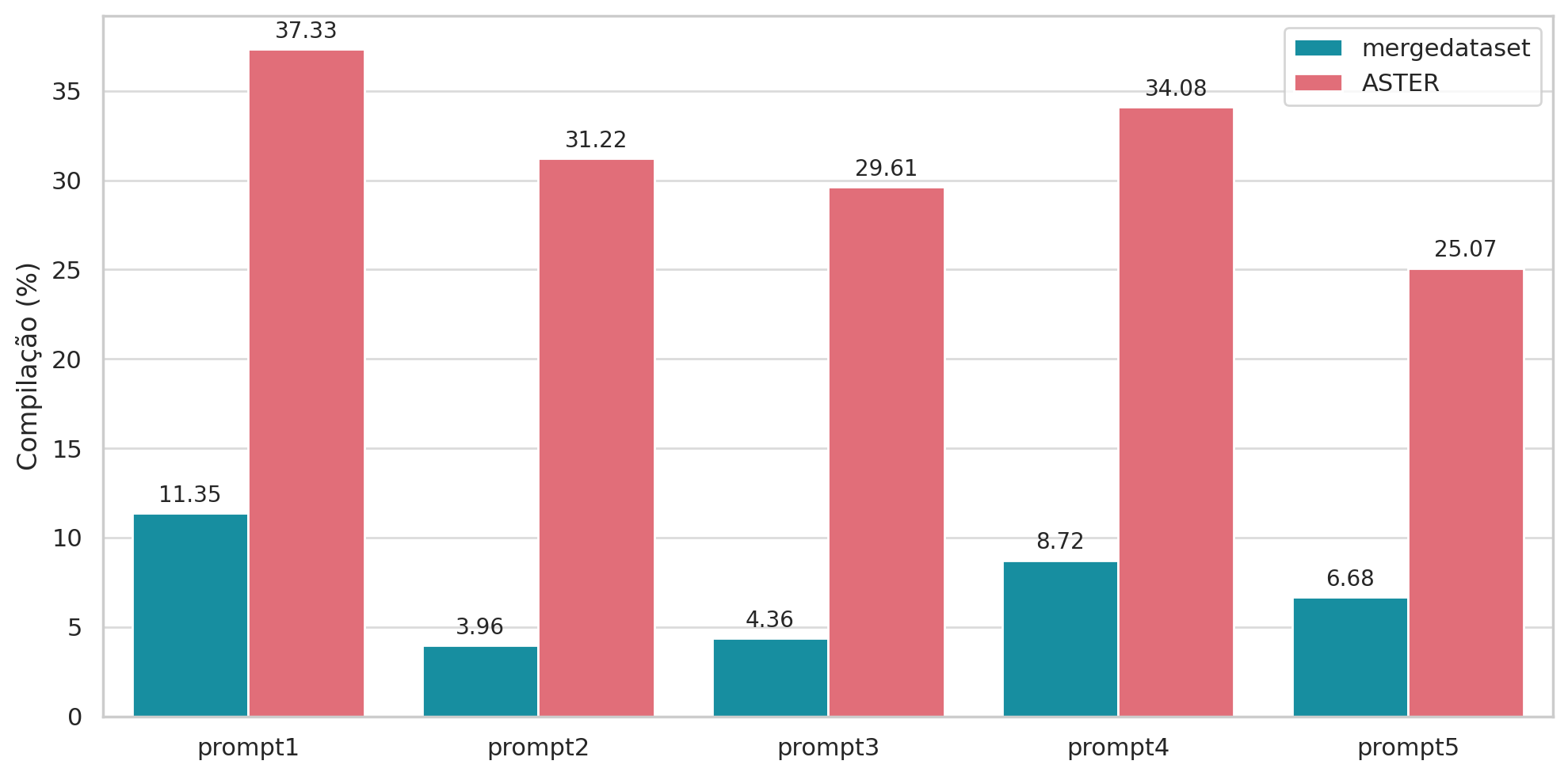}
    \caption{Comparação da taxa de compilação entre \textit{datasets} para diferentes prompts.}
    \Description{Gráfico de barras mostrando taxas de compilação da ferramenta baseada no Code Llama 70B em dois datasets distintos.}
    \label{fig:compilados_comparacao}
\end{figure}

Para avaliar o impacto da complexidade do \textit{dataset} nos resultados, comparamos a taxa de compilação dos testes gerados usando o \textit{mergedataset} e o \textit{ASTER dataset}, conforme mostrado no gráfico da Figura~\ref{fig:compilados_comparacao}. Os \textit{prompts} 1 a 5 selecionados contém as seguintes partes do contexto:

\begin{itemize}
    \item \texttt{prompt1}: (1-shot) atributos da classe, construtores e corpo do método-alvo.
    \item \texttt{prompt2}: (zero-shot) corpo do método-alvo.
    \item \texttt{prompt3}: (zero-shot) atributos da classe e corpo do método-alvo.
    \item \texttt{prompt4}: (zero-shot) construtores e corpo do método-alvo.
    \item \texttt{prompt5}: (zero-shot) atributos da classe, construtores e corpo do método-alvo.
\end{itemize}

Os resultados mostram um aumento drástico e consistente na taxa de compilação ao usar o \textit{ASTER}: o ganho relativo variou de +228.9\% a +688.9\%. 
É importante notar que este estudo focou na capacidade de compilação dos testes gerados, sem avaliar métricas de qualidade como cobertura de código ou consistência de qualidade entre múltiplas execuções dos \textit{prompts}, aspectos que poderiam complementar a análise da eficácia da ferramenta.
Esses dados ajudam a esclarecer a terceira pergunta de pesquisa:

\begin{tcolorbox}[colback=rq3bg, colframe=brown!40!black, title=RQ3]
Observamos que a taxa de compilação foi bem maior no \textit{ASTER dataset} em comparação com o \textit{mergedataset}. Isso sugere que a complexidade do código do \textit{dataset} influencia diretamente a capacidade do modelo de gerar testes compiláveis.
\end{tcolorbox}

\subsection{Tempo de Execução}
\label{sec:tempo_execucao}

\begin{table}[!htbp]
\centering
\small
\renewcommand{\arraystretch}{1.2}
\begin{tabular}{@{}p{5.5cm}l@{}}
\toprule
\textbf{Abordagem} & \textbf{Tempo de Execução} \\
\midrule
EvoSuite & 6h 7min \\
Differential EvoSuite & 1h 57min \\
\textbf{Code Llama 70B (\textit{1-shot}, temperatura 0.0)} & 27h 45min \\
\textbf{Code Llama 70B (\textit{zero-shot}, temperatura 0.0)} & 22h 58min \\
\textbf{Code Llama 70B (\textit{1-shot}, temperatura 0.7)} & 6h 5min \\
\textbf{Code Llama 70B (\textit{zero-shot}, temperatura 0.7)} & 9h 28min \\
\textbf{Code Llama 70B (união de todas as configurações experimentais)} & 132h 32min \\
\bottomrule
\end{tabular}
\caption{
Tempos de geração para o \textit{mergedataset} por ferramenta, considerando um \textit{timeout} de 300 segundos. Para a ferramenta baseada no Code Llama 70B: tempos médios de duas execuções com \textit{seeds} diferentes, 8 \textit{prompts} cada.
}
\label{tab:tempo_execucao}
\end{table}

A Tabela~\ref{tab:tempo_execucao} mostra os tempos médios de geração dos testes para cada configuração da ferramenta baseada no Code Llama 70B, além dos tempos estimados para EvoSuite e Differential EvoSuite.
A configuração mais rápida (\textit{1-shot}, temperatura 0.7) levou em média cerca de 6 horas, tempo comparável ao EvoSuite, enquanto a mais lenta (\textit{1-shot}, temperatura 0.0) ultrapassou 27 horas.

O Differential EvoSuite se destaca pela maior eficiência, detectando 6 conflitos em cerca de 1h 57min, resultando na melhor taxa de conflitos por hora (3.06 conflitos/hora).
O EvoSuite apresenta uma eficiência intermediária com 0.82 conflitos/hora, enquanto a melhor configuração individual do Code Llama 70B (\textit{zero-shot}, temperatura 0.0) alcança 0.13 conflitos/hora ao detectar 3 conflitos em 22h 58min.

A união de todas as oito configurações do Code Llama 70B resulta em um tempo total de 132h 32min.
Apesar de detectar o mesmo número de conflitos que o EvoSuite (5 conflitos), essa abordagem tem uma eficiência de apenas 0.038 conflitos/hora, sendo aproximadamente 21 vezes menos eficiente que o EvoSuite, e 80 vezes menos que o Differential EvoSuite.
Esses dados indicam que a ferramenta baseada no Code Llama 70B ainda apresenta desafios significativos em termos de eficiência e custo computacional.
\section{Ameaças à Validade}
\label{sec:ameacas-a-validade}

Este estudo apresenta limitações relevantes que podem impactar a generalização dos resultados.
O foco exclusivo no modelo Code Llama 70B pode introduzir viés específico, além do risco de contaminação dos dados de treinamento, já que alguns projetos do \textit{mergedataset} podem ter sido utilizados durante o treinamento do modelo.
Essa questão, no entanto, é válida para qualquer modelo de linguagem, especialmente os de código.
Apesar da exploração de diferentes configurações de temperatura, formatos de \textit{prompt} e \textit{seeds}, o conjunto de configurações testadas representa apenas uma pequena parcela das possibilidades, não abrangendo outras técnicas de \textit{prompt engineering}, nem outros parâmetros que poderiam influenciar os resultados.
A realização dos experimentos em uma única configuração de \textit{hardware} pode comprometer a reprodutibilidade dos resultados em outros ambientes computacionais.
Além disso, os \textit{datasets} utilizados, embora compostos por projetos Java reais, podem não refletir toda a diversidade de projetos existentes, já que se restringem a projetos de código aberto.
O foco em Java também limita a extrapolação dos resultados para outras linguagens de programação, que apresentam características sintáticas e semânticas distintas.

Outro ponto importante é o tamanho amostral reduzido (apenas 79 cenários de \textit{merge}), o que limita a significância estatística das conclusões e indica a necessidade de amostras maiores para análises mais robustas.
As métricas de avaliação adotadas concentram-se principalmente na taxa de compilação e detecção de conflitos, sem considerar aspectos como cobertura de código ou qualidade dos testes gerados.
A avaliação qualitativa dos testes também foi restrita, não abordando critérios como legibilidade e utilidade prática para desenvolvedores.
Por fim, a natureza não determinística dos LLMs pode introduzir variabilidade nos resultados que não foi completamente caracterizada, mesmo com o uso de \textit{seeds} fixas.
\section{Trabalhos Relacionados}
\label{sec:trabalhos-relacionados}

Esta seção apresenta uma análise detalhada dos trabalhos relacionados, organizados em ferramentas para detecção e resolução de conflitos semânticos em \textit{merges}, e abordagens para geração automatizada de testes utilizando LLMs. Para cada categoria, discutimos as metodologias empregadas, os resultados obtidos e as principais diferenças em relação à nossa proposta.

\paragraph{Detecção e Resolução de Conflitos Semânticos}

O problema de conflitos semânticos em \textit{merge} tem sido abordado por diferentes perspectivas na literatura. O SafeMerge~\cite{safemerge}, utilizado como \textit{baseline} no trabalho de~\citet{leuson2020}, aplica verificação composicional para detectar conflitos semânticos em \textit{merges}. No entanto, sua abordagem gera mais falsos positivos quando usamos os mesmos critérios adotados pelo SMAT para caracterizar conflitos. Especificamente, o SafeMerge apresenta uma taxa de 3.8\% de imprecisão, enquanto nossa abordagem baseada em LLMs alcançou uma taxa de 0\% de falsos positivos, demonstrando maior precisão na detecção de conflitos semânticos. Para efeito de comparação, o próprio SMAT apresenta uma taxa de 3\%, o que reforça a vantagem da nossa proposta nesse aspecto.

O trabalho de~\citet{zhang2022using} propõe a ferramenta \emph{GMERGE}, que sugere soluções automáticas para conflitos de \textit{merge}, incluindo tanto conflitos textuais quanto semânticos. Assim como nossa abordagem, \emph{GMERGE} utiliza LLMs pré-treinados e explora estratégias de \textit{k-shot} na criação de \textit{prompts}. No entanto, há diferenças importantes: \emph{GMERGE} foca em conflitos de um único projeto e sua avaliação está centrada na observação de mensagens do compilador, enquanto nosso trabalho considera conflitos semânticos identificados por meio de testes quebrados. Além disso, \emph{GMERGE} explora principalmente a viabilidade de reparar conflitos, enquanto nosso foco está na detecção.

A complementaridade entre \emph{GMERGE} e nossa abordagem é particularmente interessante: enquanto \emph{GMERGE} atua como uma ferramenta de reparo focada em conflitos que já se manifestaram através de erros de compilação, nossa proposta funciona como um ``detector preventivo'' que identifica conflitos semânticos. Esta diferença de escopo pode sugerir como trabalho futuro uma abordagem híbrida, onde nossa metodologia seria aplicada primeiro para detectar conflitos latentes, seguida por uma versão modificada do \emph{GMERGE} para resolvê-los.

\paragraph{Geração de Testes com LLMs}

No contexto da geração de testes com LLMs, destacamos quatro abordagens recentes: \emph{TESTPILOT}~\cite{testpilot}, \emph{ASTER}~\cite{aster}, \emph{CITYWALK}~\cite{citywalk} e \emph{HITS}~\cite{hits}.

O \emph{TESTPILOT} é voltado para JavaScript; seu \textit{pipeline} inclui a extração de comentários de documentação para enriquecer o contexto fornecido ao modelo, enquanto nosso \textit{pipeline} se baseia apenas em informações estruturais do código (atributos, construtores e corpo do método), sem recorrer a documentação textual.
Paradoxalmente, esta limitação pode se tornar uma vantagem no contexto de detecção de conflitos semânticos, pois força o LLM a se basear exclusivamente na estrutura e semântica do código, reduzindo a possibilidade de ser ``enganado'' por documentação inconsistente.
Além disso, o \emph{TESTPILOT} prioriza a geração de testes que não falham, buscando maximizar a cobertura, enquanto nosso foco está na detecção de conflitos semânticos em cenários de \textit{merge}.

O \emph{ASTER}, por sua vez, é uma ferramenta para geração de testes em Java e Python, que explora diferentes contextos e técnicas de \textit{mocking} para aumentar a cobertura dos testes gerados. Diferentemente da nossa abordagem, que executa a geração de testes em uma única etapa para cada cenário, o \emph{ASTER} adota um processo iterativo, refinando os testes até atingir uma cobertura desejada. Além disso, o \emph{ASTER} enfatiza a utilidade e compreensibilidade dos testes gerados, aspectos também considerados em nossa análise, mas com ênfase na capacidade de detecção de conflitos.

Já o \emph{CITYWALK} propõe um \textit{pipeline} para geração de testes em C++ que incorpora técnicas de \textit{Retrieval-Augmented Generation} (RAG). Embora nossa abordagem não utilize RAG, reconhecemos o potencial dessa técnica para trabalhos futuros, especialmente porque o conhecimento contextual externo pode ser crucial para detectar conflitos semânticos de interações complexas entre diferentes partes do sistema.
O trabalho usa diversos modelos em sua análise, enquanto nossa proposta se concentra exclusivamente no Code Llama 70B. No entanto, a ferramenta incorporada ao SMAT pode ser facilmente adaptada para usar outros modelos.
Os resultados mostram que a ferramenta gera menos erros de compilação e alcança maior cobertura com menos testes quando comparada a soluções que utilizam os modelos diretamente, reforçando o potencial dos \textit{pipelines} de LLMs na geração de testes.

Por fim,~\citet{hits} propõem uma abordagem para geração de testes de alta cobertura em métodos Java complexos.
O HITS resolve o problema de LLMs terem desempenho ruim em métodos complexos, decompondo-os em ``fatias'' de código e gerando testes para cada uma através de \textit{prompting Chain-of-Thought} (CoT), diferindo da nossa estratégia de \textit{prompting} mais simples.
O ``fatiamento'' simplifica o escopo de análise do LLM, aumentando a cobertura dos testes gerados.
O HITS supera outras abordagens baseadas em LLM e o EvoSuite em termos de cobertura de linha e \textit{branch}.
Essa abordagem é relevante para nossa proposta, pois ambas investigam a geração de testes com LLMs em métodos complexos e utilizam o EvoSuite como referência, embora com focos diferentes.
Outras distinções estão na escolha do modelo (gpt-turbo-3.5-0125 vs Code Llama 70B), nos parâmetros analisados (HITS avalia \textit{top-p} e temperatura; nossa abordagem, \textit{seed} e temperatura) e na inclusão, pelo HITS, de uma etapa para recuperação de testes não executáveis, ausente em nossa metodologia.
\section{Conclusão}
\label{sec:conclusao}

Este trabalho demonstrou que uma ferramenta baseada em modelos de linguagem, como o Code Llama 70B, representa uma alternativa promissora para a detecção de conflitos semânticos.
Os resultados superam ou se igualam a ferramentas automatizadas tradicionais, detectando cinco conflitos distintos dentre os 29 presentes no \textit{dataset}, incluindo um conflito inédito, sem introduzir falsos positivos.

Os experimentos revelaram três descobertas sobre o uso de LLMs para detecção de conflitos semânticos.
Primeiro, nenhum \textit{prompt} isolado foi capaz de identificar todos os conflitos observados: a combinação de diferentes execuções mostrou-se mais eficaz, evidenciando que diferentes configurações detectam diferentes tipos de conflitos.
Entre as configurações testadas, os \textit{prompts zero-shot} com temperatura 0.0 apresentaram melhor desempenho no \textit{mergedataset}, identificando até três conflitos em uma única execução.
Segundo, a temperatura do modelo influencia significativamente os resultados.
Nas execuções com temperatura 0.7, os \textit{prompts 1-shot} superaram os \textit{zero-shot}, reforçando a importância de explorar diferentes estratégias de configuração para otimizar a detecção.
E, em terceiro, não existe correlação evidente entre o número de testes compilados e a quantidade de conflitos identificados.
Mas, ainda assim, a taxa de compilação mais baixa no \textit{mergedataset} pode estar relacionada à complexidade dos cenários, conforme sugerido pelos experimentos no \textit{ASTER dataset}.

Os resultados apresentam implicações diretas para o desenvolvimento de ferramentas baseadas em LLMs para detecção de conflitos semânticos.
A necessidade de combinar múltiplas configurações sugere que trabalhos futuros devem focar em estratégias de combinação inteligente de diferentes configurações de \textit{prompts} e parâmetros.
A sensibilidade dos modelos às configurações de temperatura e tipo de \textit{prompt} evidencia a importância de desenvolver métodos adaptativos que ajustem automaticamente esses parâmetros conforme o contexto do código analisado.
O custo computacional significativo observado (21 vezes maior que o EvoSuite) aponta para a necessidade de pesquisas em otimização de eficiência, mantendo a eficácia na detecção.
Trabalhos futuros devem investigar métodos para reduzir esse custo, como o uso de modelos menores e mais especializados.

\textbf{Como direções futuras}, pretendemos investigar métodos para melhorar a taxa de compilação dos testes gerados, incluindo uma análise das causas das falhas de compilação e o desenvolvimento de estratégias para corrigir problemas relacionados a \textit{imports} e outras questões sintáticas.
Adicionalmente, planejamos desenvolver estratégias de ajuste de parâmetros e explorar a combinação adaptativa de \textit{prompts}.
Além disso, planejamos investigar o uso de técnicas de RAG, agentes e outros modelos de linguagem para criar ferramentas mais eficientes e eficazes na detecção de conflitos semânticos.
\section*{DISPONIBILIDADE DE ARTEFATOS}

O código-fonte da versão do SMAT utilizada neste trabalho, incluindo a integração com o Code Llama, está disponível em~\cite{smat-codellama}.
Os artefatos experimentais podem ser acessados em~\cite{online-appendix}.

\section*{AGRADECIMENTOS}
Agradecemos aos integrantes do Software Productivity Group, ao INES (Instituto Nacional de Engenharia de Software) pelo apoio disponibilizando os servidores necessários para a execução dos experimentos, e ao CNPq (projeto 465614/2014-0) pelo financiamento deste trabalho.


\clearpage

\appendix

\end{document}